\documentclass[11pt]{article}

\def\be{\begin{equation}}
\def\ee{\end{equation}}
\def\ba{\begin{eqnarray}}
\def\ea{\end{eqnarray}}
\def\nn{\nonumber}
\def\lb{\label}
\def\bb{\bibitem}
\def\A{{\cal A}}
\begin{document}

\begin{titlepage}

\date{1 February 2021}

\title{
\begin{flushright}\begin{small}    LAPTH-044/20
\end{small} \end{flushright} \vspace{1cm}
The gravimagnetic dipole}

\author{G\'erard Cl\'ement\thanks{Email: gclement@lapth.cnrs.fr} \\ \\
{\small LAPTh, Universit\'e Savoie Mont Blanc, CNRS,} \\ {\small 9 chemin de Bellevue,
BP 110, F-74941 Annecy-le-Vieux cedex, France}}

\maketitle

\begin{abstract}
We investigate a previously constructed stationary solution of the vacuum Einstein equations, which represents a system of two non-extreme black holes with equal masses and opposite NUT charges, connected by a Misner string with tension. For large separations, the inverse square law force measured by this tension is attractive or repulsive, according to the relative values of the masses and NUT charges. For small separations, the force is always repulsive, so that the system cannot collapse to a single black hole. For given values of the black hole masses and NUT charges, there is a unique configuration such that the Misner string is tensionless. This behaves asymptotically as the Kerr solution, but can be overspinning while remaining free from a ring singularity, thus evading the usual black hole uniqueness theorems. All double black hole and string configurations satisfy a generalized first law of black hole mechanics where the two black holes and the Misner string are treated on an equal footing.
\end{abstract}
\end{titlepage}
\setcounter{page}{2}

\section{Introduction}

A recurring question in general relativity is that of the existence of regular, stationary, asymptotically flat double black hole systems, as solutions either to the vacuum Einstein equations or to the Einstein-Maxwell equations. With the exception of linear Majumdar-Papapetrou \cite{MP} superpositions of extreme Reissner-Nordstr\"om black holes, these generically can be balanced only if there is a line of conical singularities (cosmic string) connecting the two black holes. Recently we have investigated some systems of two extreme black holes carrying opposite gravimagnetic or NUT charges, and shown that the parameters can be fine tuned so that the Misner string between the two black holes is tensionless \cite{dimagn,baldimagn}. Misner strings are usually considered to be unphysical, but we have shown that they are transparent to geodesic motion, and argued that they do not lead to observable violations of causality \cite{GC15}.

Non-extreme double black hole solutions of the Einstein-Maxwell equations have been discussed in \cite{Emparan:2001bb}. The superposition of two Kerr-NUT solutions was analyzed in \cite{Tomimatsu:1981bc}, where it was shown that balance could be achieved in the absence of both cosmic string and Misner string. It was not clear however whether the two constituent sources were regular black holes, or whether the solution was free from a ring singularity. A similar problem was later discussed in \cite{letelier98a}.

After the double Schwarzschild solution, where balance is clearly impossible, the simplest asymptotically flat stationary double black hole solution is the double Schwarzschild-NUT vacuum solution. Such a solution was constructed in \cite{manko2009} as a non-linear superposition of two sources with equal masses and opposite NUT charges. The main purpose of the authors of \cite{manko2009} was to show that their solution reduced to the Kerr black hole when the distance between the two sources was appropriately chosen. In this paper, we analyze in more detail this solution, show that it is free from a ring singularity in a large parameter domain, and determine the two-dimensional surface in the three-dimensional parameter domain such that the system is weakly balanced, in the sense that the Misner string connecting the two constituent black holes is tensionless.

The solution is recalled in the next section. The geometrical structure which it describes is identified in section 3, where the condition for the absence of a Kerr-like ring singularity is given, and the horizon characteristics are determined. The properties of the string connecting the two black holes are discussed in section 4. In section 5, we evaluate the horizon and string Komar masses and angular momenta, and show that the dipole system satisfies a generalized first law of black hole mechanics where the two black holes and the Misner string are treated on an equal footing. Our conclusions are given in the final section.

\setcounter{equation}{0}
\section{The solution}

The non-linear superposition of two NUT solutions (of the vacuum Einstein equations)with equal masses $m$
and opposite NUT charges $\nu$, separated by a distance $2k$ ($k \ge m >0$), leads to the Ernst potential,
constructed by Sibgatullin's method \cite{sibga84} in \cite{manko2009} :
 \be
{\cal E} = \frac{A-B}{A+B},
 \ee
with
 \ba
A &=& \left[(m^2+\nu^2)(k^2-m^2)(k^2-m^2-\nu^2) - 2m^2k^2\nu^2\right](R_+-R_-)(r_+-r_-) \nn\\
&& - \alpha_+\alpha_-\left[2(m^2+\nu^2)(k^2-m^2)(R_+R_- + r_+r_-)\right. \nn\\
&& \left. + (2m^4 + (m^2+\nu^2)(k^2-m^2))(R_++R_-)(r_++r_-)\right] \nn\\
&& - 2imk\nu d\left[(\alpha_+-\alpha_-)(R_+r_+ - R_-r_-) - \right. \nn\\
&& \left. (\alpha_++\alpha_-)(R_+r_- - R_-r_+)\right], \\
B &=& 4d\left\{m\alpha_+\alpha_-\left[(m^2-d)(R_++R_-) - (m^2+d)(r_++r_-)\right]\right. \nn\\
&& \left. + ik\nu\left[\alpha_-(m^2-d)(R_+-R_-) - \alpha_+(m^2+d)(r_+-r_-)\right]\right\},
 \ea
where
 \be
R_\pm = \sqrt{\rho^2+(z\pm\alpha_+)^2}, \quad r_\pm = \sqrt{\rho^2+(z\pm\alpha_-)^2},
 \ee
$\rho$ and $z$ being the Weyl coordinates, and
 \be\lb{dalpha}
\alpha_\pm = \sqrt{m^2+k^2-\nu^2 \pm 2d}, \quad d = \sqrt{m^2k^2 + \nu^2(k^2-m^2)}.
 \ee

The corresponding metric is
 \be
ds^2 = - f(dt-\omega d\varphi)^2 + f^{-1}\left[e^{2\gamma}(d\rho^2+dz^2) + \rho^2d\varphi^2\right],
 \ee
with
 \be\label{metfunct}
f = \frac{A\bar{A} - B\bar{B}}{(A+B)(\bar{A}+\bar{B})}, \quad
e^{2\gamma} = \frac{A\bar{A} - B\bar{B}}{64d^4\alpha_+^2\alpha_-^2R_+R_-r_+r_-}, \quad
\omega = - \frac{4{\rm Im}[G(\bar{A}+\bar{B})]}{A\bar{A} - B\bar{B}},
 \ee
 \ba
G &=& - d[d^2+m^2(m^2+2ik\nu)]\left[(\alpha_+-\alpha_-)(R_+r_+ - R_-r_-) - (\alpha_++\alpha_-)(R_+r_- - R_-r_+)\right] \nn\\
&& + 2m^2d^2\left[(\alpha_++\alpha_-)(R_+r_+ - R_-r_-) - (\alpha_+-\alpha_-)(R_+r_- - R_-r_+)\right] \nn\\
&& - m\alpha_+\alpha_-(d^2+m^4)(R_++R_-)(r_++r_-) + m\left[kd^2(k+4i\nu) - (2k^2-m^2)(m^2+\nu^2)^2\right. \nn\\
&& \left. + k^2\nu^4\right](R_+-R_-)(r_+-r_-) - 2m\alpha_+\alpha_-(m^2+\nu^2)(k^2-m^2)(R_+R_- + r_+r_-) \nn\\
&& - 2dz\left\{\alpha_-(m^2-d)\left[m\alpha_+(R_++R_-) + ik\nu(R_+-R_-)\right] \right. \nn\\
&& \left. - \alpha_+(m^2+d)\left[m\alpha_-(r_++r_-) + ik\nu(r_+-r_-)\right]\right\} \nn\\
&& + 2d\alpha_+\alpha_-(2m^2+ik\nu)\left[m^2(R_++R_--r_+-r_-) - d(R_++R_-+r_++r_-)\right] \nn\\
&& - 2md\left[d^2-m^4-ik\nu(2m^2+ik\nu)\right]\left[\alpha_-(R_+-R_-) - \alpha_+(r_+-r_-)\right] \nn\\
&& + 2md^2(m^2-k^2+\nu^2-2ik\nu)\left[\alpha_-(R_+-R_-) + \alpha_+(r_+-r_-)\right].
 \ea
This metric is asymptotically flat, with total mass $M = 2m$ and total angular momentum $J = 2k\nu$.

The Ernst potential and the metric, given above in terms of the three parameters $m$, $k$, and $\nu$,  can be reexpressed in terms of the three parameters $\alpha_+$,
$\alpha_-$, and $m$. The definitions (\ref{dalpha}) can be inverted to give
 \be
k^2-\nu^2 = \alpha_\pm^2\mp 2d - m^2, \quad k^2\nu^2 = (d\pm m^2)^2 - m^2\alpha_\pm^2 .
 \ee
Other useful identities are
 \ba\label{ids}
&& 4d = \alpha_+^2 - \alpha_-^2 = 4m^2\sigma\delta, \quad (d \pm m^2 \pm m\alpha_\pm) = m^2(\sigma+1)(\delta\pm 1), \nn\\
&& [(d-m^2)\alpha_+ + (d+m^2)\alpha_-] = 2m^3\delta(\sigma^2-1), \nn\\
&& [(d+m^2)\alpha_+ + (d-m^2)\alpha_- + 4md] = 2m^3\delta(\sigma+1)^2, \nn\\
&& k^2\nu^2 = m^4(\sigma^2-1)(\delta^2-1),
 \ea
where
 \be
\sigma \equiv \frac{\alpha_++\alpha_-}{2m}, \quad \delta \equiv \frac{\alpha_+-\alpha_-}{2m}.
 \ee

Using these, the three functions $A$, $B$ and $G$ can be reexpressed in terms of $\alpha_+$, $\alpha_-$, and $m$ as:
 \ba
A &=& \frac12\left[(d-m^2)^2\alpha_+^2 + (d+m^2)^2\alpha_-^2\right](R_+-R_-)(r_+-r_-) \nn\\
&& - \alpha_+\alpha_-\left[2(d^2-m^4)(R_+R_- + r_+r_-) + (d^2+m^4)(R_++R_-)(r_++r_-)\right] \nn\\
&& - 2imk\nu d\left[(\alpha_+-\alpha_-)(R_+r_+ - R_-r_-) - (\alpha_++\alpha_-)(R_+r_- - R_-r_+)\right], \nn\\
B &=& - 4d\left\{(d-m^2)\alpha_-\left[m\alpha_+(R_++R_-) + ik\nu(R_+-R_-)\right]\right. \nn\\
&& \left. + (d+m^2)\alpha_+\left[m\alpha_-(r_++r_-) + ik\nu(r_+-r-)\right]\right\} \nn\\
G &=& d[- (d-m^2)^2\alpha_+ + (d+m^2)^2\alpha_- - 2ik\nu m^2(\alpha_+-\alpha_-)](R_+r_+ - R_-r_-) \nn\\
&& + d[(d-m^2)^2\alpha_+ + (d+m^2)^2\alpha_- + 2ik\nu m^2(\alpha_++\alpha_-)](R_+r_- - R_-r_+) \nn\\
&& - m(d^2+m^4)\alpha_+\alpha_-(R_++R_-)(r_++r_-) \nn\\
&& + \frac{m}2\left[(d-m^2)^2\alpha_+^2 + (d+m^2)^2\alpha_-^2 + 8ik\nu d^2\right](R_+-R_-)(r_+-r_-) \nn\\
&& - 2m(d^2-m^4)\alpha_+\alpha_-(R_+R_- + r_+r_-) \\
&& - 2d(d-m^2)\alpha_-\left[(\alpha_+ +2m-z)(m\alpha_+ + ik\nu)R_+ - (\alpha_+ -2m+z)(m\alpha_+ - ik\nu)R_-\right] \nn\\
&& - 2d(d+m^2)\alpha_+\left[(\alpha_- +2m-z)(m\alpha_- + ik\nu)r_+ - (\alpha_- -2m+z)(m\alpha_- - ik\nu)r_-\right] .\nn
 \ea

\setcounter{equation}{0}
\section{Structure}

For $\nu=0$, $d=km$ and $\alpha_\pm = k\pm m$ are real, and the static metric describes a system of two black holes of mass $m$
connected by a cosmic string of length $2(k-m)$. The two black holes coalesce when $k=m$ to form a single Schwarzschild black hole.

The situation is less simple when $\nu\neq0$, and depends on the domain of values of $k$.
For $k$ large enough, $d$ and $\alpha_\pm$
are again real and the system consists of two black holes connected by a spinning cosmic string (Misner string). On the $z$ axis,
these correspond to three rods, two symmetrical rods $\alpha_-<z<\alpha_+$ and $-\alpha_+<z<-\alpha_-$ corresponding to the two black hole horizons, and the central rod $-\alpha_-<z<\alpha_-$ corresponding to the Misner string. We have checked that on the axis outside the rods
($|z|>\alpha_+$), $G$, which is a polynomial of second order in $z$, vanishes identically, so that the axis condition $\omega(0,z)=0$ is satisfied. Therefore the solution is asymptotically flat. The stationary metric now depends on the three parameters $m$, $k$ and $\nu$, or $m$, $\sigma$ and $\delta$. These are related to the equal black hole masses $M_{H_\pm}$ and black hole angular momenta $J_{H_\pm}$, which shall be evaluated in section 5, and the black hole separation, related to the string length $2\alpha_-=2m(\sigma-\delta)$. The other string characteristics, mass, angular momentum, and tension, are dynamically fixed so that the system of two black holes and string is in stationary equilibrium.

It was shown in \cite{manko2009} that for $k=m$ the solution describes the Kerr black hole (a single rod) provided $|\nu|<2m$ (and presumably the extreme black hole
if $|\nu|=2m$). In this case, $\alpha_+$ is real and $\alpha_-$ imaginary. Clearly, there will be an intermediate domain of values
$k>m$ with $\alpha_+$ real and $\alpha_-$ imaginary, and it is not clear what will the solution describe in this case. We have found that $\alpha_\pm$ are real and non-zero provided
 \be\lb{kpm}
k > k_+ \;\; {\rm or} \;\; k < k_-, \quad k_\pm(m,\nu) = \sqrt{m^2+2\nu^2} \pm |\nu|.
 \ee
The reality of the parameters $\alpha_\pm$ will be assumed in the following.
We will show in section 4 that the solution for $k = k_+(m,\nu)$  is actually singular for $\nu\neq0$. It follows that when the separation $2k$ between the two NUTty black holes with fixed $m$ and $\nu$ is decreased, a singular configuration with $k = k_+(m,\nu)$ is reached {\em before} the Kerr configuration $k = m$.

\subsection{Absence of ring singularity}
The reality condition (\ref{kpm}) is satisfied in two disjoint parameter domains.
We now prove that the domain $k>k_+$ is free from an equatorial ring singularity, which is always present in the domain $k<k_-$ .

The equatorial plane is $z=0$, so that $R_+=R_-=R$,
and $r_+=r_-=r$, leading to
 \ba
A &=& - 2\alpha_+\alpha_-[(d^2-m^4)(R+r)^2 + 4m^4Rr], \nn\\
B &=& -8dm\alpha_+\alpha_-[(d-m^2)R + (d+m^2)r].
 \ea
For $\rho\to\infty$,
 \be
A + B \simeq -8\alpha_+\alpha_-d^2\rho^2,
 \ee
while, for $\rho=0$,
 \be
A + B = - 8m^6\alpha_+\alpha_-\delta^2(\sigma+1)^2(\sigma^2-1)
 \ee
so that a necessary condition for the absence of a zero of $A+B$ is $\sigma>1$, implying $\delta>1$ in view of the last equation (\ref{ids}), and so $d = m^2\sigma\delta > m^2$, equivalent to $k>m$. Conversely, it follows from
 \be
d^2-m^4=(m^2+\nu^2)(k^2-m^2)
 \ee
that $k>m$ implies $d>m^2$, so that $A<0$ and $B\le0$, and therefore $A+B$ cannot vanish, proving that the equatorial ring singularity is absent.

So the necessary and sufficient condition for the absence of an equatorial ring singularity is $k>m$. This is clearly satisfied in the sector $k>k_+$, but violated in the sector $0<k<k_-$, where $k^2 < m^2 - 2k\nu$. In the following we will always assume $k>k_+$.

\subsection{Horizons}

On the upper horizon ($\rho=0$, $\alpha_-<z<\alpha_+$),
 \be
R_\pm = \alpha_+ \pm z, \quad r_\pm = z \pm \alpha_- ,
 \ee
leading to
 \ba
A_H &=& -8d\alpha_-\left\{(d+m^2)[(d-m^2)\alpha_+ + (d+m^2)z] + ik\nu m(\alpha_+^2-z^2)\right\}, \\
B_H &=& -8d\alpha_-\left\{m\alpha_+[(d-m^2)\alpha_+ + (d+m^2)z] + ik\nu[(d+m^2)\alpha_+ + (d-m^2)z] \right\}. \nn
 \ea

We know that $e^{2\gamma}$ and $\omega$ are constant over each rod.
So their values for a given rod can be obtained by evaluating the corresponding expressions (\ref{metfunct})
at the axis point $z=0$ (whether or not it belongs to the rod). For the horizon rod, noting $A_H(0,0) = A_0$, etc.,
we obtain:
 \ba
A_0 &=& -8d\alpha_+\alpha_-\left[d^2-m^4 + ik\nu m\alpha_+\right], \nn\\
B_0 &=& -8d\alpha_+\alpha_-\left[m\alpha_+(d-m^2) + ik\nu (d+m^2)\right],\\
G_0 &=& -4d(d+m^2)\alpha_+\alpha_-\left[(\alpha_+-2m)(d+m\alpha_++m^2) + ik\nu(\alpha_++2m)\right], \nn
 \ea
leading to the horizon characteristics
 \be
e^{2\gamma_H} = - \left(\frac{k\nu m}{d\alpha_+}\right)^2, \quad \Omega_H = \omega_H^{-1} = \frac{k\nu m}{2(d+m^2)(d+m\alpha_++m^2)}.
 \ee
The surface gravity $\kappa_H = \sqrt{|e^{-2\gamma_H}|}\,|\Omega_H|$ is
 \be
\kappa_H = \frac{d\alpha_+}{2(d+m^2)(d+m\alpha_++m^2)} = \frac{\sigma\delta(\sigma+\delta)}
{2m(1+\sigma\delta)(\sigma+1)(\delta+1)},
 \ee
from which one can derive the horizon area ${\cal A}_H = 2\pi L_H/\kappa_H$ (where
$L_H = \alpha_+-\alpha_- = 2m\delta$ is the length of the horizon rod).

\setcounter{equation}{0}
\section{The string}

The two co-rotating black holes, corresponding to the rods ($\rho=0$, $\alpha_-<z<\alpha_+$) and ($\rho=0$, $-\alpha_+<z<-\alpha_-$), are connected by a finite length spinning cosmic string
corresponding to the rod ($\rho=0$, $-\alpha_-<z<\alpha_-$),
On this string,
 \be
R_\pm = \alpha_+ \pm z, \quad r_\pm = \alpha_- \pm z,
 \ee
leading to
 \ba
A_S &=& - 2\alpha_+\alpha_-[d^2(\alpha_++\alpha_-)^2 - m^4(\alpha_+-\alpha_-)^2], \nn\\
&& + 2[d(\alpha_++\alpha_-) - m^2(\alpha_+-\alpha_-)]^2z^2 - 32ik\nu md^2z \nn\\
B_S &=& - 8dm\alpha_+\alpha_-[d(\alpha_++\alpha_-) - m^2(\alpha_+-\alpha_-)] \nn\\
&& - 8ik\nu d[d(\alpha_++\alpha_-) + m^2(\alpha_+-\alpha_-)]z .
 \ea
Evaluating these at the string center $z=0$ (where they are real), together with ${\rm Im}\,G$,
 \be
{\rm Im}\,G_0 = -4k\nu d\alpha_+\alpha_-[d(\alpha_++\alpha_-) - m^2(\alpha_+-\alpha_-)], \nn
 \ee
we obtain the string characteristics
 \ba
e^{2\gamma_S} &=& \left(\frac{[d(\alpha_++\alpha_-) - m^2(\alpha_+-\alpha_-)]^2}{4d^2\alpha_+\alpha_-}\right)^2
\;=\; \left(\frac{(\sigma^2 - 1)^2}{\sigma^2(\sigma^2-\delta^2)}\right)^2, \\
\omega_S &=& -8k\nu d\frac{d(\alpha_++\alpha_-+4m) + m^2(\alpha_+-\alpha_-)}{[d(\alpha_++\alpha_-) - m^2(\alpha_+-\alpha_-)]^2}
\;=\; -\frac{4k\nu\sigma}{m(\sigma-1)^2}. \nn
 \ea
From these we can derive the string area and surface gravity,
 \be
\A_S = 4\pi\alpha_-|\omega_S|e^{\gamma_S} = \frac{16\pi k\nu(\sigma+1)^2}{\sigma(\sigma+\delta)}, \quad
\kappa_S = \frac{4\pi m(\sigma-\delta)}{\A_S}.
 \ee

These characteristics are simply related to the string tension per unit length, which is
$(1-e^{-\gamma_S})/4$, and the string spin, $-\omega_S/4$.
As discussed in \cite{taleful}, this spin measures the gravimagnetic flow along the Misner string connecting the two horizons carrying
the effective NUT charges $\mp\omega_S/4$ (generically different from the ``bare'' NUT charges $\pm\nu$). And the string tension
measures the force \cite{letelier98b} which is necessary to balance the gravitational forces between the two sources (black holes), attractive
between the two equal masses and repulsive between the opposite NUT charges, a negative string tension ($e^{\gamma_S}<1$) corresponding
to a net attractive force, and a positive string tension ($e^{\gamma_S}>1$) corresponding to a net repulsive force.

For a large separation between the two black holes ($k \gg (m,\,\nu)$, $\sigma\gg1$), we find $\sigma \simeq k/m$, $\delta \simeq \mu/m$
($\mu^2 = m^2 + \nu^2$). The string length $2\alpha_-$ is of the order of $2k$, while the string tension and spin go to
 \be
\frac{1-e^{-\gamma_S}}4 \simeq - \frac{m^2-\nu^2}{(2k)^2}, \qquad - \frac{\omega_S}4 \simeq \nu .
 \ee
In this quasi-linear newtonian regime, the effective and bare NUT charges coincide, while the net force is simply the superposition of the inverse
square law gravitational and gravimagnetic forces

On the other hand, when for fixed $m$ and $\nu$ the parameter $k$ is progressively decreased to its lowest possible limit $k_+(m,\nu)$,
corresponding to $\alpha_-\to0$ ($\delta \to \sigma$), $e^{\gamma_S}$ diverges while the string length goes to zero,
signalling a singularity at $\rho=z=0$, unless $\sigma$ simultaneously goes to $1$, i.e. $k$ goes to $m$, which
from the condition $k>k_+$ is possible only for $\nu=0$. Otherwise, whatever the relative values of the mass and NUT parameters $m$ and $\nu$, the
regime becomes highly non-linear when the limit $k\to k_+$ is approached, leading to an increasing repulsion which will prevent the two
black holes from coalescing, no matter what external force is exerted.

In the absence of an external force, the stationary double black hole system is balanced provided
the string parameters are constrained by $e^{\gamma_S} = 1$, which is achieved by
 \be\lb{bal}
\delta^2 = 2 - \frac1{\sigma^2}.
 \ee
The regular balanced configurations depend therefore only on two parameters, which can be chosen for instance as the total mass $M=2m$
and the dimensionless parameter $\sigma\ge1$.
Recalling that the total angular momentum is $J \equiv Ma = 2k\nu$, for these balanced configurations the ratio
 \be\lb{aoverM}
\frac{|a|}M = \frac{k|\nu|}{2m^2} = \frac12\left(\sigma - \sigma^{-1}\right)
 \ee
can take any real non-negative value. Using this relation and the relation
 \be
k^2 - \nu^2 = m^2(\sigma^2 + 1 - \sigma^{-2}),
 \ee
one derives the relation between the ratio $\nu/m$ and the parameter $\sigma$,
 \be
\nu^4 + (\sigma^2 + 1 - \sigma^{-2})m^2\nu^2 - (\sigma - \sigma^{-1})^2m^4 = 0.
 \ee
For $\sigma\ge1$, $\nu/m$ is bounded above by $1$. More precisely, when $\sigma$ varies from $1$ to infinity,
$\nu/m$ varies from $0$ to $1$.

For fixed mass $M=2m$ and large $\sigma$, one finds
 \be
\nu \simeq m\sigma(1-2\sigma^{-2}), \qquad k \simeq m\sigma(1+\sigma^{-2}).
 \ee
The string rod length increases as
$L_S \simeq M\sigma$, while the horizon rod lengths go to the finite value $L_H \simeq M\sqrt2$.
The horizon area and surface gravity go to finite limits, and the horizon angular velocity goes to zero,
 \be
\A_H \simeq 4\pi\,\frac{1+\sqrt2}{\sqrt2}\,M^2, \quad \kappa_H \simeq \frac1{1+\sqrt2}\,M^{-1},
\quad \Omega_H \simeq \frac1{\sqrt2(1+\sqrt2)}\,(M\sigma)^{-1},
 \ee
in spite of the fact that the total angular momentum diverges as (\ref{aoverM}). Of course, this
large angular momentum is due to the gravimagnetic dipole moment $L_S \simeq 2k\nu$.

In the opposite limit $\sigma\to 1$, one can take $\nu/m \simeq 2a/M$ as independent dimensionless parameter, in terms of which
 \be
\sigma \simeq 1 + \frac\nu{2m} + \frac{5\nu^2}{8m^2}, \qquad k \simeq m + \nu.
 \ee
The horizon rod lengths are now $L_H \simeq M$, while the string rod contracts to a length $L_S \simeq 2a^2/M$. The values
of the {\em total} horizon area, surface gravity, and horizon angular velocity
 \be
2\A_H \simeq 16\pi M^2, \quad \kappa_H \simeq \frac1{4M}, \quad \Omega_H \simeq \frac{a}{4M^2}
 \ee
coincide in lowest order with those of a slowly rotating Kerr black hole with the same parameters
$M$ and $a$.

\setcounter{equation}{0}
\section{Komar observables and first law}

Now we compute the horizon and string Komar masses and angular momenta. The Tomimatsu formulas \cite{Tomimatsu:1983qc} for these are
 \be\lb{MnJn}
M_n = \frac{\omega_n}4\,[\chi(z_{n+}) - \chi(z_{n-})], \quad J_n = \frac{\omega_n}2\left(M_n - \frac{z_{n+} - z_{n-}}2\right),
 \ee
where $\chi = - {\rm Im}\cal E$\footnote{The Ernst potential used in \cite{manko2009} and in the present paper
is the complex conjugate of that used in our previous papers.}, and $z_{n\pm}$ are the upper and lower ends of
the rod, $\pm\alpha_\pm$ and $\pm\alpha_\mp$ for the upper and lower horizon rods, and $\alpha_-$ and $-\alpha_-$
for the string rod. At these ends,
 \ba
A(\pm\alpha_+) &=& -16d^2(d+m^2)\alpha_+\alpha_-, \nn\\
B(\pm\alpha_+) &=& -16d^2\alpha_+\alpha_-(m\alpha_+ \pm ik\nu), \nn\\
A(\pm\alpha_-) &=& -8d\alpha_-\left\{(d+m^2)[(d-m^2)\alpha_+ + (d+m^2)\alpha_-] \pm ik\nu4md\right\}, \nn\\
B(\pm\alpha_-) &=& -8d\alpha_- \left\{m\alpha_+[(d-m^2)\alpha_+ + (d+m^2)\alpha_-] \right.\nn\\
&& \left. \pm ik\nu[(d+m^2)\alpha_+ + (d-m^2)\alpha_-]\right\}.
 \ea
Using these and the identities (\ref{ids}), we obtain,
 \ba
\chi(\pm\alpha_+) &=& \pm \frac{k\nu}{m^2(\sigma+1)(\delta+1)}, \nn\\
\chi(\pm\alpha_-) &=& \pm \frac{k\nu(\sigma-1)}{m^2(\sigma+1)^2(\delta+1)},
 \ea
Together with the rod angular velocities evaluated in the previous section, these lead to the horizon
and string Komar masses and angular momenta
 \ba
M_{H_\pm} &=& m\,\frac{\sigma\delta+1}{\sigma+1}, \quad J_{H_\pm} = - k\nu\,\frac{\sigma\delta+1}{\sigma^2-1}, \nn\\
M_S &=& - 2m\,\frac{\sigma(\delta-1)}{\sigma+1}, \quad J_S = 2k\nu\,\frac{\sigma(\sigma+\delta)}{\sigma^2-1}.
 \ea
It is easy to check that these add up to the total mass and angular momentum,
 \be
2M_H + M_S = 2m, \quad 2J_H + J_S = 2k\nu.
 \ee

For a large separation between the two black holes ($k \gg (m,\,\nu)$, $\sigma\gg1$),
the Komar angular momentum is mainly carried by the string,
while the Komar masses of the two horizons and of the string are of the same order but opposite signs,
$M_{H_\pm} \simeq M\delta/2$ and $M_S \simeq -M(\delta-1)$ (with $\delta \simeq \sqrt2$ for a balanced configuration).
Conversely, when the configuration is balanced and $\nu\to0$, the Komar mass is mainly carried by the black holes,
the respective horizon and string Komar angular momenta being of the order of $M^2$ but nearly compensating each other.

By construction, each rod satisfies its own Smarr relation (this is a simple consequence of the Tomimatsu formulas
(\ref{MnJn})), so that the global Smarr relation \cite{smarrnut}
 \be
M = 2\left(2\Omega_{H}J_{H} + \frac{\kappa_{H}}{4\pi} \A_{H}\right)  +
2\Omega_{S}J_{S} + \frac{\kappa_{S}}{4\pi} \A_{S}
 \ee
is trivially satisfied. Note that all the Killing horizons -- the two black hole horizons and the Misner string -- contribute equally to this global Smarr law.

One may surmise that a similar global first law of mechanics
 \be\lb{firstdinut}
dM = \sum_{i}\left(\Omega_{H_i}\,dJ_{H_i} + \frac{\kappa_{H_i}}{8\pi}\,d\A_{H_i}\right)  + \Omega_{S}\,dJ_{S} + \frac{\kappa_{S}}{8\pi}\,d\A_{S}
 \ee
($i=1,2$)
should hold for all systems of two black holes connected by a Misner string\footnote{This will be proved in \cite{talefirst}.}. In the present case, careful accounting shows that the differential first law (\ref{firstdinut}) indeed holds for independent variations of the two dimensionless parameters $\sigma$ and $\delta$ (it is trivially satisfied for variations of the scale $m$), whether the system is balanced or unbalanced. It should be stressed that, contrary to the Smarr relations, the first law is not satisfied independently by the various interacting components -- horizons and string -- but only by the composite system.

The form (\ref{firstdinut}) of the first law for two NUTty black holes should be compared with that recently proposed \cite{krtous}-\cite{gregory} for collinear systems of NUTless black holes connected by cosmic strings, which involves the string tensions. For instance, the first law for a system of two aligned Kerr black holes reads \cite{manko2020a}
 \be\lb{firstdikerr}
dM = \sum_{i}\left(\Omega_{H_i}\,dJ_{H_i} + \frac{\kappa_{H_i}}{8\pi}\,d\A_{H_i}\right)
+ \lambda_S\,d\mu_S,
 \ee
with $\mu_S$ the tension of the string connecting the two black holes and $\lambda_S$ the conjugate string thermodynamic length \cite{gregory}. While in the present case the string tension does not explicitly occur in the first law (\ref{firstdinut}), it is related to the Misner string characteristics by
 \be
\mu_S = \frac{1-e^{-\gamma_S}}4, \quad e^{-\gamma_S} = -\frac{\kappa_S}{\Omega_S}.
 \ee
In the NUTless limit $\delta\to1$ ($\nu\to0$), the Misner string intensive quantities $M_S$, $J_S$ and $\A_S$ all vanish, while the extensive quantities $\Omega_S$ and $\kappa_S$ both diverge, their ratio being related to the string tension by the above relation. We find that in this limit the string contribution to (\ref{firstdinut}) does not vanish, but goes to the limit
 \be
\Omega_{S}\,dJ_{S} + \frac{\kappa_{S}}{8\pi}\,d\A_{S} = \frac{m\,d\sigma}{\sigma(\sigma+1)} = \lambda_S\,d\mu_S ,
 \ee
where the thermodynamic length of the string between the two Schwarzschild black holes is
 \be
\lambda_S = \frac{2m(\sigma+1)(\sigma-1)^2}{\sigma^2} = 2\alpha_-\,e^{\gamma_S},
 \ee
in accordance with the result of \cite{krtous} (where the distance $R$ between the centers of the two black-hole rods is related to our parameters by $R=2m\sigma)$.

\setcounter{equation}{0}
\section{Discussion}

We have discussed the properties of an exact asymptotically flat stationary solution to vacuum gravity, describing a system of two non-extreme black holes with equal masses and opposite NUT charges, connected by a Misner string with tension. The solution is free from a ring singularity in a certain parameter domain. For large separations, the force measured by the string tension is attractive or repulsive, depending on the relative values of the masses and NUT charges. For small separations, the force is always repulsive, so that the
system cannot collapse to a single black hole. For given values of the black hole masses and NUT charges, the distance between the two black holes (the string length) can always be adjusted so that the string tension vanishes and the system is balanced. The existence of a Misner string, on which the (usually enforced) axis condition $\omega(0,z) = 0$ is relaxed, is necessary for this balance to be possible.

This condition being relaxed, the usual black hole uniqueness theorems no longer apply. The observable charges of the dihole solution (total mass and angular momentum) take the same values as those of the Kerr solution, while the overall topology and geometry are different. For small black hole separations (short dipoles), the angular-momentum-to-mass ratio is small, and the global horizon characteristics (horizon
angular velocity, temperature and entropy) are roughly the same as those of a slowly rotating Kerr black hole with the same parameters $M$ and $a=J/M$. On the other hand, for large black hole separations (long dipoles), the ratio $|a|/M$ becomes arbitrarily large, corresponding to regular overspinning configurations.

Finally, we have evaluated the horizon and string Komar masses and angular momenta and other mechanical observables, and showed that the dipole system satisfies a generalized first law of black hole mechanics (\ref{firstdinut}), to which all the Killing horizons -- the two black hole horizons and the Misner string -- contribute equally.

\section*{Acknowledgments}
I warmly thank Dmitry Gal'tsov for a critical reading of the manuscript and useful suggestions.

\end{document}